\documentclass[twoside,twocolumn,english,aps,prb,superscriptaddress]{revtex4-1}
\usepackage[latin9]{inputenc}
\usepackage{amsmath}
\usepackage{amssymb}
\usepackage{graphicx}
\usepackage{esint}
\usepackage{float}
\usepackage{times}
\usepackage[colorlinks,linkcolor=blue,anchorcolor=blue,citecolor=blue]{hyperref} 
\makeatletter

\usepackage{dcolumn}
\usepackage{bm}
\usepackage{xcolor}

\usepackage{babel}
\makeatother
\usepackage{babel}

\begin{document}

\title{Topological states and topological phase transition in Cu$_{2}$SnS$_{3}$ and Cu$_{2}$SnSe$_{3}$}
\author{Liqin Zhou}
\affiliation{Beijing National Laboratory for Condensed Matter Physics and Institute of physics, Chinese academy of sciences, Beijing 100190, China}
\affiliation{University of Chinese academy of sciences, Beijing 100049, China}

\author{Yuting Qian}
\affiliation{Beijing National Laboratory for Condensed Matter Physics and Institute of physics, Chinese academy of sciences, Beijing 100190, China}
\affiliation{University of Chinese academy of sciences, Beijing 100049, China}

\author{Changming Yue}
\affiliation{Department of Physics, University of Fribourg, 1700 Fribourg, Switzerland}

\author{Zhong Fang}
\affiliation{Beijing National Laboratory for Condensed Matter Physics and Institute of physics, Chinese academy of sciences, Beijing 100190, China}
\affiliation{University of Chinese academy of sciences, Beijing 100049, China}

\author{Wei Zhang}
\email{zhangw721@163.com} 
\affiliation{Fujian Provincial Key Laboratory of Quantum Manipulation and New Energy Materials, College of Physics and Energy, Fujian Normal University, Fuzhou 350117, China}
\affiliation{Fujian Provincial Collaborative Innovation Center for Advanced High-Field Superconducting Materials and Engineering, Fuzhou 350117, China}

\author{Chen Fang}
\email{cfang@iphy.ac.cn} 
\affiliation{Beijing National Laboratory for Condensed Matter Physics and Institute of physics, Chinese academy of sciences, Beijing 100190, China}
\affiliation{University of Chinese academy of sciences, Beijing 100049, China}

\author{Hongming Weng}
\email{hmweng@iphy.ac.cn} 
\affiliation{Beijing National Laboratory for Condensed Matter Physics and Institute of physics, Chinese academy of sciences, Beijing 100190, China}
\affiliation{University of Chinese academy of sciences, Beijing 100049, China}
\affiliation{Songshan Lake Materials Laboratory, Dongguan, Guangdong 523808, China}

\begin{abstract}
Based on the first-principles calculations within local density approximation and model analysis, we propose that the iso-structural compounds Cu$_{2}$SnS$_{3}$ and Cu$_{2}$SnSe$_{3}$ are both the simplest nodal line semimetals with only one nodal line in their crystal momentum space when spin-orbit coupling (SOC) is ignored. The including of SOC drives Cu$_{2}$SnS$_{3}$ into a Weyl semimetal (WSM) state with only two pairs of Weyl nodes, the minimum number required for WSM with time reversal symmetry. In contrast, SOC leads Cu$_{2}$SnSe$_{3}$ to strong topological insulator (TI) state. This difference can be well understood as there is a topological phase transition (TPT). In it, the Weyl nodes are driven by tunable SOC and annihilate in a mirror plane, resulting in a TI. This TPT, together with the evolution of Weyl nodes, the changing of mirror Chern numbers of mirror plane and the $Z_2$ indices protected by time-reversal symmetry has been demonstrated by the calculation of Cu$_{2}$SnS$_{3-x}$Se$_{x}$ within virtual crystal approximation and an effective $k\cdot p$ model analysis. Though our first-principles calculations have overestimated the topological states in both compounds, we believe that the theoretical demonstration of controlling TPT and evolution of Weyl nodes will stimulate further efforts in exploring them. 

\end{abstract}

\maketitle
\section{Introduction}
After nearly fifteen years development, the classification of topological electronic bands and their topological materials have been quite well developed.~\cite{PhysRevB.78.195125, song2018quantitative,zhang2019catalogue,vergniory2019complete,tang2019comprehensive, xu2020high, peng2021topological, PhysRevX.7.041069, PhysRevB.103.245127} The gapped states have been classified both with internal and spatial symmetries. The internal symmetries include time-reversal symmetry, chiral (sub-lattice) symmetry and particle-hole symmetry and the spatial symmetries include the crystalline symmetries in all four types of magnetic space groups. The Chern insulator, integer quantum anomalous Hall insulator, topological insulator (TI), topological crystalline insulator (TCI) as well as topological superconductor belong to these classifications. For the metals, the topological classification has been done mainly according to the nodal points close to Fermi energy. According to the degeneracy, topological charge and distribution of these nodes, there have been Weyl semimetal (WSM), Dirac semimetal (DSM), topological nodal-line semimetal (TNLS) and multiple-degeneracy nodal point semimetal.~\cite{RevModPhys.82.3045,RevModPhys.83.1057,wan2011topological,PhysRevLett.107.127205,PhysRevLett.108.140405,PhysRevB.85.155118,yang2014classification,weng2015weyl,PhysRevB.92.045108,armitage2018weyl,  doi:10.1080/00018732.2015.1068524, weng2016topological, 10.1093/nsr/nwx066, doi:10.7566/JPSJ.87.041001,fang2016topological}. 
Topological semimetal phases can be viewed as the intermediate states in the process of the topological phase transition (TPT) between different topological phases, such as the normal insulator (NI) to TI, which has been systematically studied by Murakami {\it et al}. ~\cite{murakami2007phase,PhysRevB.76.205304,PhysRevB.78.165313,murakami2011gap,PhysRevB.89.235315,Murakamie1602680,hu2021evolution} In inversion-symmetric systems, the conduction band and valence band gradually approach each other in the phase transition process, and the band gap closes at time-reversal invariant momenta (TRIM) only, where a four-fold degenerate Dirac nodes appeared. The intermediate state of the phase transition is a DSM phase, but as a critical point it is unstable and easy to be destroyed. On the other hand, for inversion-asymmetric systems, the band gap will close at certain $k$-point away from TRIM and at least two pairs of Weyl nodes with opposite chirality will emerge as constrained by time-reversal symmetry (TRS) and no-go theorem. This intermediate state of the phase transition is a WSM phase. The Weyl nodes are separated in reciprocal space and they should appear and disappear in pair when tuning one or more parameters in the Hamiltonian properly. In this sense, the intermediate WSM phase can not be destroyed immediately and it is relatively robust against perturbation, facilitating the material realization.

In this work, we have theoretically investigated two iso-structural compounds Cu$_{2}$SnS$_{3}$ and Cu$_{2}$SnSe$_{3}$. We find they are very suitable to study the TPT from a WSM to a TI. When spin-orbit coupling (SOC) is ignored, both of them are nodal line semimetal with only one nodal ring around the Fermi energy lying in one mirror plane. When SOC is considered, Cu$_{2}$SnS$_{3}$ becomes a WSM with two pairs of Weyl nodes, while Cu$_{2}$SnSe$_{3}$ is a strong TI. The evolution and annihilation of Weyl nodes in this iso-structural and iso-electronic family compounds can be demonstrated by systematically tuning the effective SOC. To do this, we employ the virtual crystal approximation (VCA) method to simulate the different Se doping concentration of Cu$_{2}$SnS$_{3-x}$Se$_{x}$. Since there are no topologically nontrivial symmetry-based indicators in their space group \emph{Imm}2 (No. 44) to directly judge their topological classification,~\cite{po2017symmetry,song2018quantitative,PhysRevX.8.031069} we characterize their topological states by calculating the mirror Chern number (MCN) for two mutually perpendicular mirror planes and $Z_2$ indices according to the Wilson loop method.~\cite{yu2011equivalent, PhysRevB.95.075146} To reveal the mechanism of TPT, an effective $k\cdot p$ model has been constructed and analyzed according to the representations of the bands forming the nodal ring and Weyl nodes. In the following, we firstly introduce the calculation method, and then discuss the topological states of Cu$_{2}$SnS$_{3}$ and Cu$_{2}$SnSe$_{3}$ without and with SOC, respectively. Finally, the TPT from WSM in Cu$_{2}$SnS$_{3}$ to TI of Cu$_{2}$SnSe$_{3}$ has been investigated systematically.

\section{method}

The density functional theory (DFT) calculation of the electronic structures for Cu$_{2}$SnS$_{3}$ and Cu$_{2}$SnSe$_{3}$  are performed by using the Vienna {\it ab initio} simulation package (VASP)\cite{KRESSE199615}. The generalized gradient approximation (GGA) with the Perdew-Burke-Ernzerhof (PBE) functional are selected to describe the exchange-correlation energy\cite{blochl1994projector,perdew1996generalized}. The cutoff energy for plane-wave basis is set to 520 eV and the reciprocal space is sampled by 11$\times$11$\times$11 $\Gamma$-centered \emph{k} mesh. To further calculate the topological properties of Cu$_{2}$SnS$_{3}$ and Cu$_{2}$SnSe$_{3}$ such as surface states and MCN, we have constructed the tight-binding model with the maximally-localized Wannier functions (MLWF)\cite{mostofi2008wannier90} generated for Cu 3\emph{d}, Sn 5\emph{s+p} and S (Se) 3\emph{p} (4\emph{p}) orbitals. The surface states and Fermi arcs are calculated by using the WannierTools package\cite{wu2018wanniertools}, which based on the surface Green's function method. 
In order to study the TPT process between Cu$_{2}$SnS$_{3}$ and Cu$_{2}$SnSe$_{3}$, we use the VCA method (suppose different proportions of S and Se atoms occupy simultaneously the same atomic sites) to calculate the band structures with different S:Se ratios. Similary, we mix linearly the above tight-binding Hamiltonian based on Wannier functions of the two parent compounds to obtain the Hamiltonian of the doped one, which is found to be efficient in further determining the phase transition critical point and the evolution of the Weyl points.

\section{Results and discussions}
The crystal structure and Brillouin zone (BZ) of Cu$_{2}$SnS$_{3}$ and Cu$_{2}$SnSe$_{3}$ are shown in Fig. \ref{fig1}(a) and (b). They belong to the same space group \emph{Imm}2 (No. 44), which includes two mirror reflection symmetries \emph{M}$_{x}$ and \emph{M}$_{y}$ perpendicular to $x$ and $y$ axis, respectively, one twofold rotation symmetry $C_{2z}$ along $z$-axis. It has time reversal symmetry (TRS) $T$  but no inversion symmetry, which means the minimum number of Weyl nodes is four with two pairs in opposite chirality.

\subsection{Cu$_{2}$SnS$_{3}$} 

\begin{figure}[htbp!]
\centering
\includegraphics[scale=0.36]{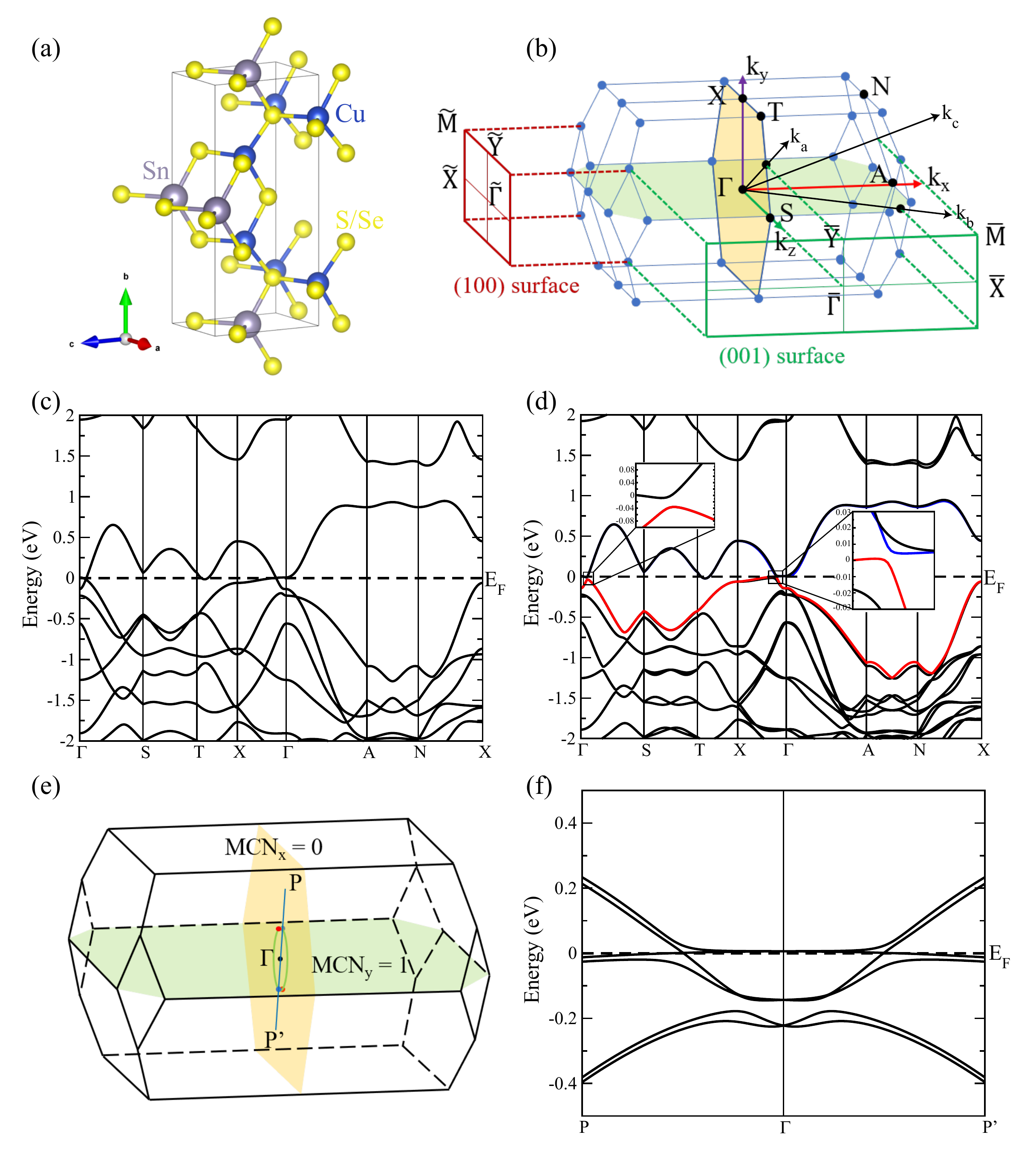}
\caption{(a) The crystal structure of Cu$_{2}$SnS$_{3}$(Se$_{3}$). The blue, grey and yellow balls represent Cu, Sn and S (Se) atoms, respectively. (b) The bulk BZ and the projected surface BZ for (001) and (100) surfaces. The light yellow and light green planes represent \emph{M}$_{x}$ and \emph{M}$_{y}$ planes, respectively. (c) and (d) The band structures of Cu$_{2}$SnS$_{3}$ without and with SOC, respectively. (e) 3D schematic diagram of the nodal-ring (in the absence of SOC) and Weyl points (with SOC) in the BZ. The green line represents the nodal-ring. The blue and red dots denote the Weyl points with opposite chirality. (f) The band along the path passing through two Weyl points with the same chirality related with time-reversal symmetry. The path P to P' is shown in (e).\label{fig1}}
\end{figure}

The band structure of Cu$_{2}$SnS$_{3}$ calculated without SOC is shown in Fig. \ref{fig1}(c). The two mirror reflection symmetries \emph{M}$_{x}$ and \emph{M}$_{y}$ are represented by the colored planes in Fig. \ref{fig1}(b). We can find clearly that the band-crossing between the highest valence band and the lowest conduction band along the $\Gamma$-S and X-$\Gamma$ directions. According to the representation analysis of the symmetric operations, the conduction band and valence band forming the crossing node on path $\Gamma$-S have opposite eigenvalues of $M_x$ and $C_{2z}$ operators but the same eigenvalues of $M_y$ operator. This means that the band-crossing is protected by the \emph{M}$_{x}$ and $C_{2z}$ symmetries. The crossing point on the path X-$\Gamma$ is also protected by the \emph{M}$_{x}$ symmetry. By searching the entire BZ, we find that these band-crossing points form a closed nodal-ring around $\Gamma$ on the $M_x$ plane as shown in Fig. \ref{fig1}(e). Therefore, without SOC Cu$_{2}$SnS$_{3}$ is the simplest TNLS with only one nodal ring.

When SOC is taken into account, the crossing points on the above mentioned nodal ring are fully gapped. The band structure near the Fermi level is plotted in Fig.~\ref{fig1}(d). However, there are two pairs of Weyl points created at the generic momenta on the $k_z$ = 0 plane. They are symmetric about the $M_x$ and $M_y$ planes. They also respect the $C_{2z}$ rotation symmetry and TRS $T$. It is noted that $k_z$ = 0 plane is invariant under the joint operation $C_{2z}* T$, which results in 0 or $\pi$ Berry phase for any loop in this plane.~\cite{peng2021topological, hu2021evolution} 
The positions and chiralities of Weyl nodes are shown in Fig.~\ref{fig1}(e). The bands along the $k$ path which connects a pair of Weyl points with the same chirality to $\Gamma$ point have been plotted in Fig.~\ref{fig1}(f). The energy of Weyl points is very close to Fermi level, being about 1.4 meV above it. 

\begin{figure}[htbp!]
\centering
\includegraphics[scale=0.1]{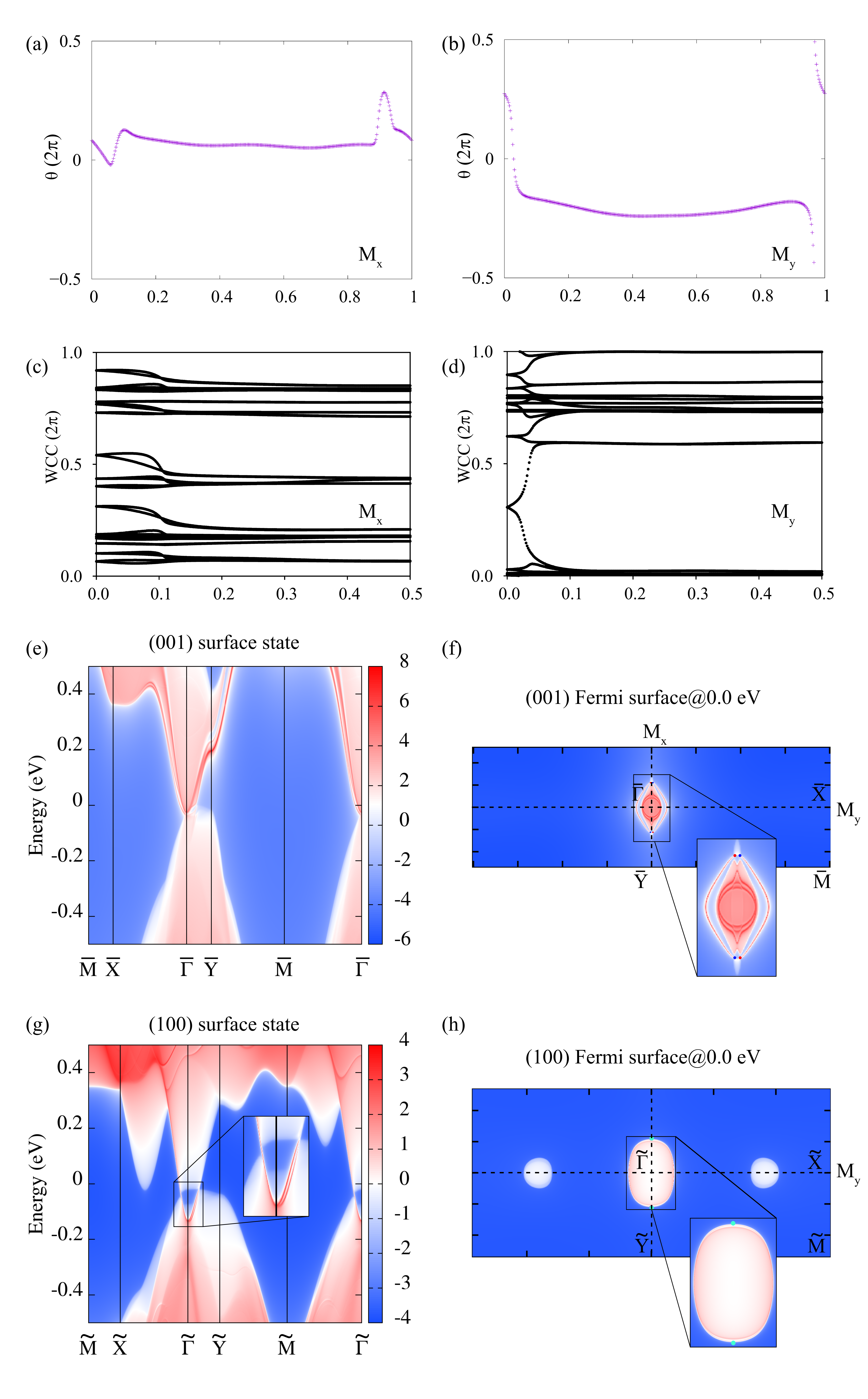}
\caption{The calculated topological properties of Cu$_{2}$SnS$_{3}$. (a) and (b) The flow chart of the average position of the Wannier centers for occupied bands with mirror eigenvalue \emph{+i} in the $M_x$ (a) and $M_y$ (b) planes. (c) and (d) The flow chart of the Wannier centers for all occupied bands in the \emph{M}$_{x}$ (c) and \emph{M}$_{y}$  (d) planes across half of the reciprocal lattice vector. (e) The surface states of (001) surface and (f) the corresponding Fermi surface; (g) The surface states of (100) surface and (h) the corresponding Fermi surface. The red and blue dots are the projections of opposite chiral Weyl points, respectively. The cyan dots are the superposition of two projected Weyl points with opposite chirality. \label{fig2}}
\end{figure}

It is noted that there is no topological indicator~\cite{po2017symmetry,song2018quantitative,PhysRevX.8.031069} that can be used to determine the topological classification in space group No. 44. Furthermore, spatial inversion symmetry is also absent so that Fu-Kane parity formula~\cite{kane2005z,fu2007topological,PhysRevB.76.045302} is not applicable. To determine its topological phase, we take the Wilson loop method to calculate the $Z_2$ invariant protected by TRS $T$ and the MCNs of mirror planes.~\cite{weng2015weyl,yu2011equivalent}
The MCN calculations for the two mirror planes are plotted in Fig.~\ref{fig2}(a) and (b). They clearly show that MCN = 0 for $M_x$ plane while MCN = 1 for $M_y$ plane. To verify the results, we further calculate the flow of Wannier centers of all occupied states along half of the reciprocal lattice vector in the $M_x$ and $M_y$ planes, which can give out $Z_2$ invariant protected by TRS. As shown in Fig.~\ref{fig2}(c) and (d), $Z_2$ is 0 for $k_x$ = 0 plane while 1 for $k_y$ = 0 plane, which is consistent with the MCN results. 
That the existence of Weyl node between two non-parallel mirror planes with different MCNs has been firstly pointed out and demonstrated in TaAs~\cite{weng2015weyl} and their influence on the pattern of Fermi arcs has also been discussed and studied experimentally in TaAs.~\cite{PhysRevX.5.031013} Therefore, the net topological charge of the Weyl nodes in one of the four blocks divided by these two mirror planes should be an odd number. In addition, $k_z=0$ plane is invariant under the joint operation of $C_2*T$. Thus, there must be Weyl nodes in this plane, which is essentially the same as the constraint of in-plane Weyl nodes in inversion-symmetric magnetic space group with odd $Z_4$ invariant and joint $C_2*T$ symmetry.~\cite{peng2021topological, hu2021evolution} 

The surface states and Fermi arcs of Cu$_{2}$SnS$_{3}$ on (100) and (001) projected surfaces are plotted in Fig.~\ref{fig2}(e)-(h). On the (001) surface, four Weyl points are all projected onto the surface separately. There are two Fermi arcs connecting the two pairs of Weyl points, respectively, which is clearly observed in the enlarged illustration of Fig. ~\ref{fig2}(f). $\bar{\Gamma}$ - $\bar{Y}$ and $\bar{\Gamma}$ - $\bar{X}$ line are the projection of $M_x$ and $M_y$ plane, respectively. Along the $\bar{\Gamma}$ - $\bar{X}$ line we can find a cross where the Fermi arc runs though it, which comes from MCN = 1 for the $M_y$ plane. There is no Fermi arc crossing the $\bar{\Gamma}$ - $\bar{Y}$ line since MCN = 0 for the $M_x$ plane. It is noted that if the Weyl nodes are off the $k_z=0$ plane, the number of Weyl nodes will be doubled and two Weyl nodes of the same chirality will be superposed on each other when projected onto (001) surface. The number of Fermi arcs connecting each projection should be two. The different MCNs for $M_x$ and $M_y$ planes limit that there must have odd number of Fermi arcs crossing $\bar{\Gamma}$ - $\bar{X}$ and even number of Fermi arcs crossing $\bar{\Gamma}$ - $\bar{Y}$. Therefore, there is no way to satisfy all these constraints if assuming the Weyl nodes were off the $k_z$ = 0 plane.

On (100) surface, two opposite chiral Weyl nodes are projected to the same point on the $\tilde{\Gamma}$ - $\tilde{Y}$ line as shown in Fig.~\ref{fig2}(h). Therefore, each projective point should be connected by two or zero Fermi arcs. Here, $\tilde{\Gamma}$ - $\tilde{X}$ line is the projection of $M_y$ plane. We can still find that one Fermi arc sticking close to the bulk state crosses this line, which is consistent with MCN = 1 for $M_y$ plane. In Fig.~\ref{fig2}(g), the projected bulk Weyl points form solid Dirac cones with continuous eigen energies along $\tilde{\Gamma}$ - $\tilde{Y}$. The surface states form an empty Dirac cone and have its Dirac node at $\tilde{\Gamma}$. Along $\tilde{\Gamma}$ - $\tilde{Y}$, both of its two branches merge into the solid Dirac cone where the bulk Weyl nodes are projected. Along $\tilde{\Gamma}$ - $\tilde{X}$, there is only one branch connecting the bulk conduction bands and the other one merges into the valence states, which is consistent with MCN = 1 for $M_y$.

\subsection{Cu$_{2}$SnSe$_{3}$} 

\begin{figure}[htbp!]
\centering
\includegraphics[scale=0.26]{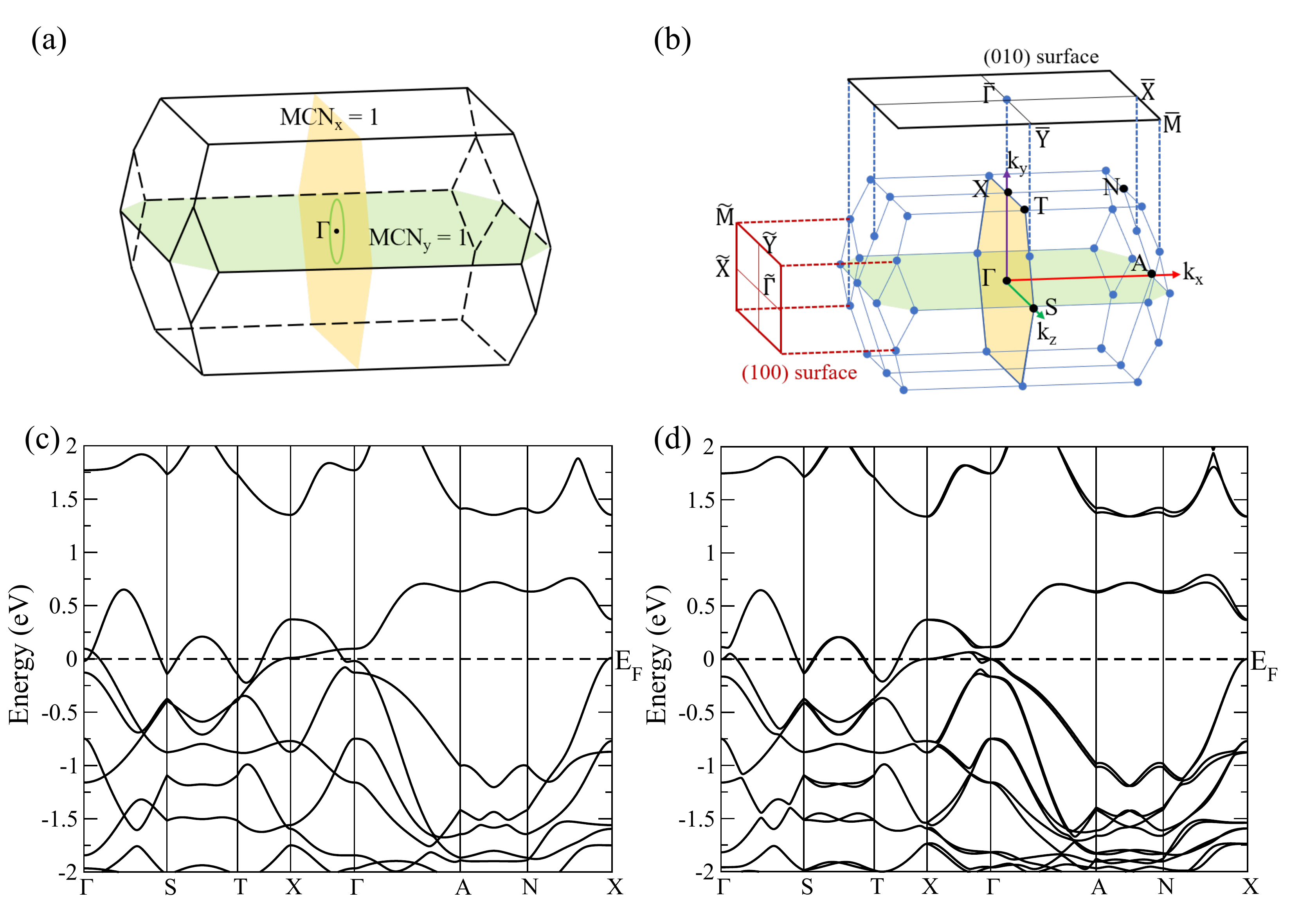}
\caption{(a) 3D schematic diagram of the nodal-ring for Cu$_{2}$SnSe$_{3}$ in the absence of SOC. (b) The bulk BZ and the projected surface BZ for (010) and (100) surfaces. (c) and (d) The band structures of Cu$_{2}$SnSe$_{3}$ without and with SOC, respectively.\label{fig3}}
\end{figure}

\begin{figure}[htbp!]
\centering
\includegraphics[scale=0.27]{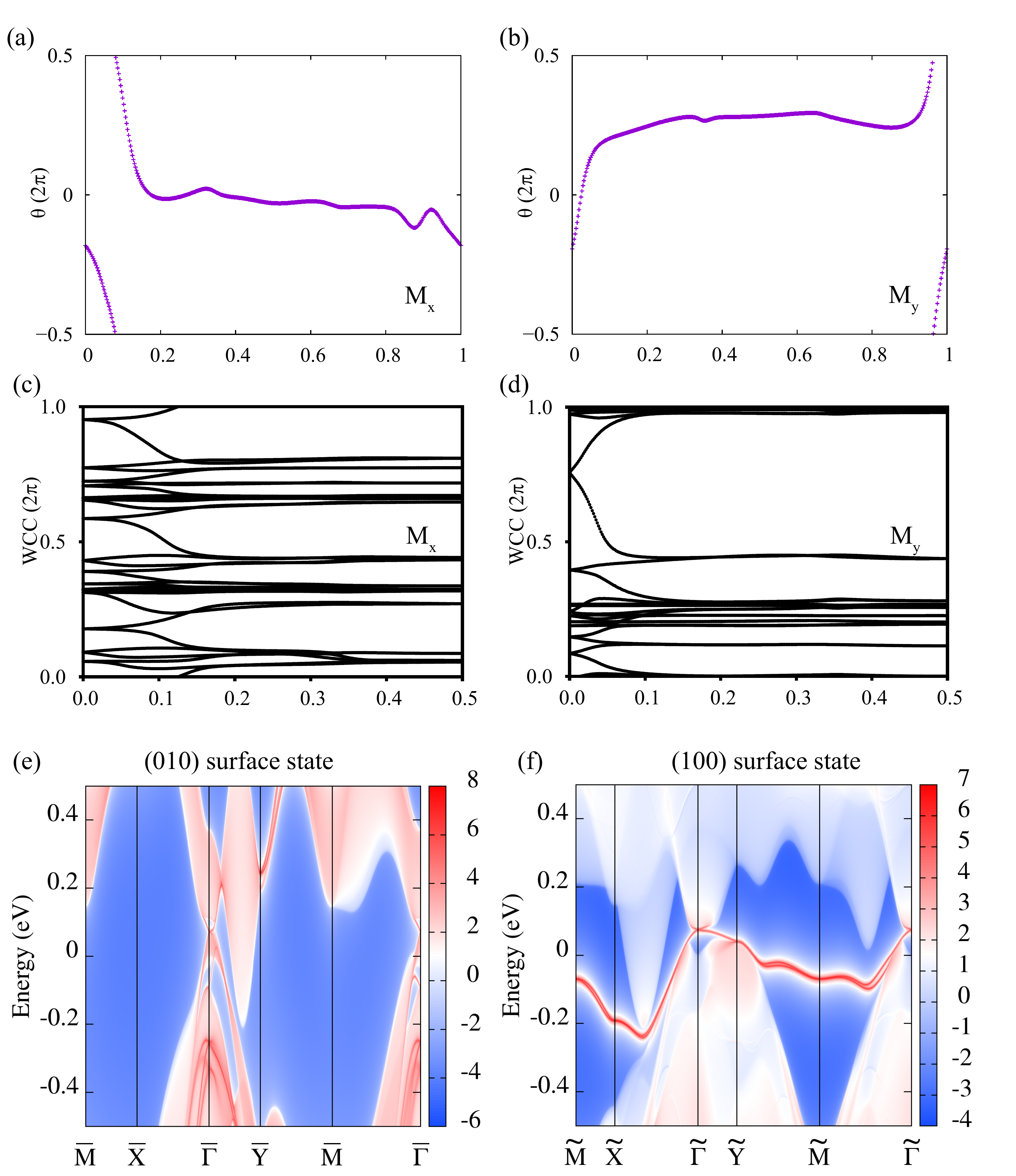}
\caption{The calculation results of the topological properties of Cu$_{2}$SnSe$_{3}$. (a) and (b) The flow chart of the average position of the Wannier centers obtained by Wilson-loop calculation for bands with mirror eigenvalue \emph{i} in the $M_x$ (a) and $M_y$ (b) planes.
(c) and (d) The flow chart of the Wannier centers of all occupied states in the \emph{M}$_{x}$ (c) and \emph{M}$_{y}$ (d) planes along half of the reciprocal lattice vector. (e) and (f) The surface states of (010) projected surface and (100) projected surface, respectively. \label{fig4}}
\end{figure}

\begin{figure*}[htbp!]
\centering
\includegraphics[width=1.0\textwidth]{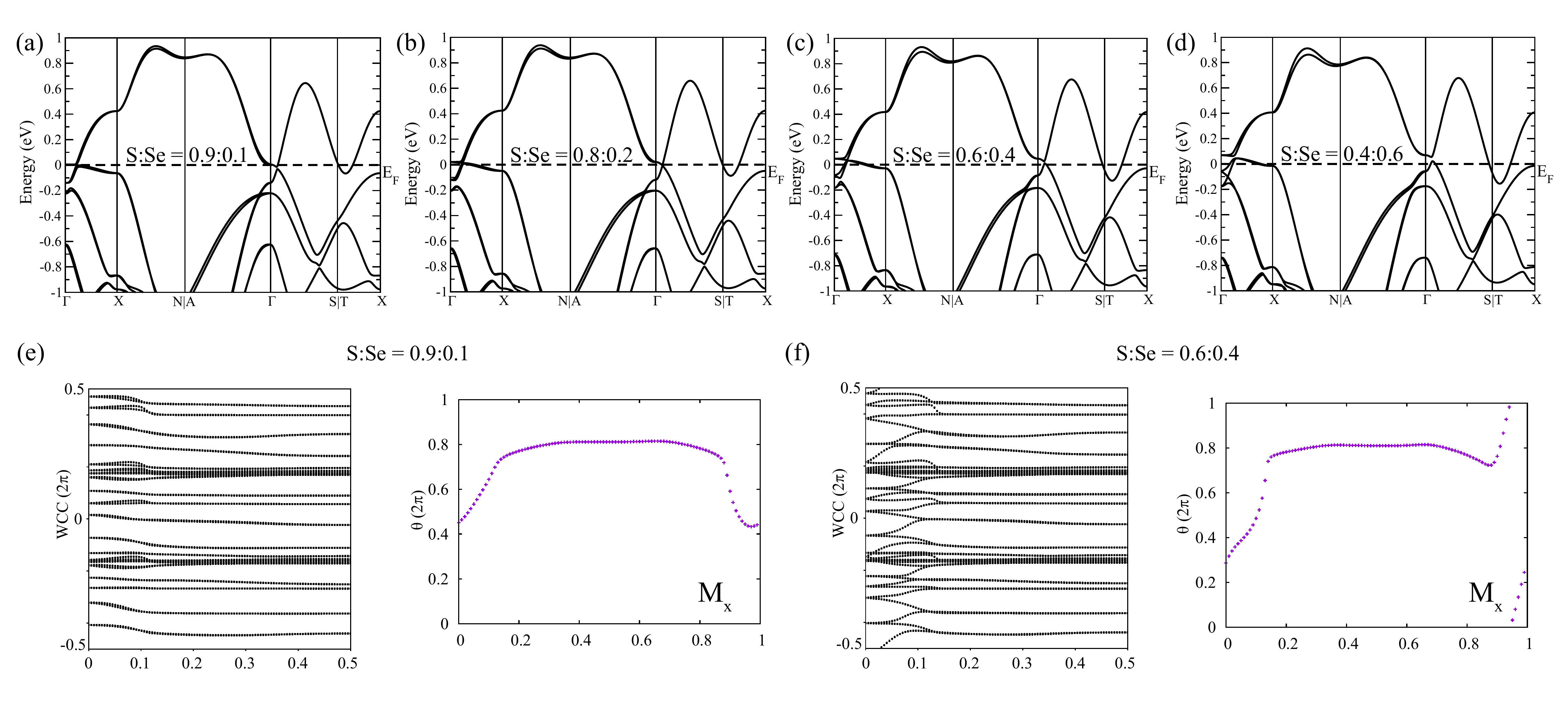}
\caption{(a)-(d) Band structures calculated by VCA method with different value of S:Se ratio. (e)-(f) The flow of Wannier centers and mirror Chern numbers of $M_x$ plane with different S:Se ratio values. \label{fig5}}
\end{figure*}

The band structures of Cu$_{2}$SnSe$_{3}$ without and with SOC are shown in Fig. \ref{fig3}(c) and (d), respectively. In the absence of SOC, Cu$_{2}$SnSe$_{3}$ is also a nodal-line semimetal with only one nodal ring centering at $\Gamma$ in the $M_x$ plane.
However, when SOC is taken into account, the band structure of Cu$_{2}$SnSe$_{3}$ is fully gapped at each $k$ point along the nodal line. It can be looked as an insulator although there is no global gap in the whole BZ. In order to determine whether it is a topologically nontrivial insulator, we further obtain the MCNs for $M_x$ and $M_y$ planes as shown in Fig. \ref{fig4}(a) and (b). It is obvious that MCN = 1 for both of $M_x$ and $M_y$ planes, consistent with the $Z_2$ invariant calculation shown in Fig. \ref{fig4}(c) and (d). Thus, Cu$_{2}$SnSe$_{3}$ might be a WSM with even number of Weyl nodes in one quarter of the BZ divided by the $M_x$ and $M_y$ planes, or a strong TI with $Z_2$ indices (1;000). We have found that the former situation is possible in another family member compound Cu$_{2}$GeSe$_{3}$ as shown in the Appendix. The present compound Cu$_{2}$SnSe$_{3}$ is the later case. The most typical feature of a TI is the appearance of odd number of Dirac cones on their surfaces. We further calculate the (010) and (100) surface states of Cu$_{2}$SnSe$_{3}$ as shown in Fig. \ref{fig4}(e) and (f). There is one Dirac cone at $\tilde{\Gamma}$ point on either (010) or (100) surface and the two branches of the Dirac cone connecting the valence and conduction bands, respectively.

\section{Topological phase transition}
As the materials of a family with the same space group, Cu$_{2}$SnS$_{3}$ and Cu$_{2}$SnSe$_{3}$ are both topological nodal ring semimetals when SOC is not taken into account, but they are obviously different in band topology when SOC is considered. It is intriguing to know about the mechanism underlying this difference. Therefore, we're going to explore the process of TPT between them continuously from a WSM to a TI by doping Se into Cu$_{2}$SnS$_{3}$, through which the strength of SOC can be tuned. 

We use the virtual crystal approximation (VCA) method to calculate the bands of Cu$_{2}$SnS$_{3-x}$Se$_{x}$ to simulate Se doping effect as shown in Fig. \ref{fig5}(a)-(d). The change in lattice constants is linearly scaled between Cu$_{2}$SnS$_{3}$ and Cu$_{2}$SnSe$_{3}$ with the doping concentration. It can be seen that the band inversion between the valence band and conduction band around $\Gamma$ keeps existing as Se doping ratio increases, and the spin splitting in these bands also increases due to enhanced SOC. This indicates the SOC is tunable.
In order to accurately determine where the phase transition has occurred, we calculated the MCN and $Z_2$ for $M_x$ plane in different doping cases. We find that when S:Se = 0.9:0.1, MCN and $Z_2$ on $M_x$ plane are both zero. When S:Se = 0.6:0.4, both of MCN and $Z_2$ on $M_x$ plane become one, as shown in Fig. \ref{fig5}(e)-(f), indicating that TPT has occurred around this point.

\begin{figure}[H]
\centering
\includegraphics[scale=0.157]{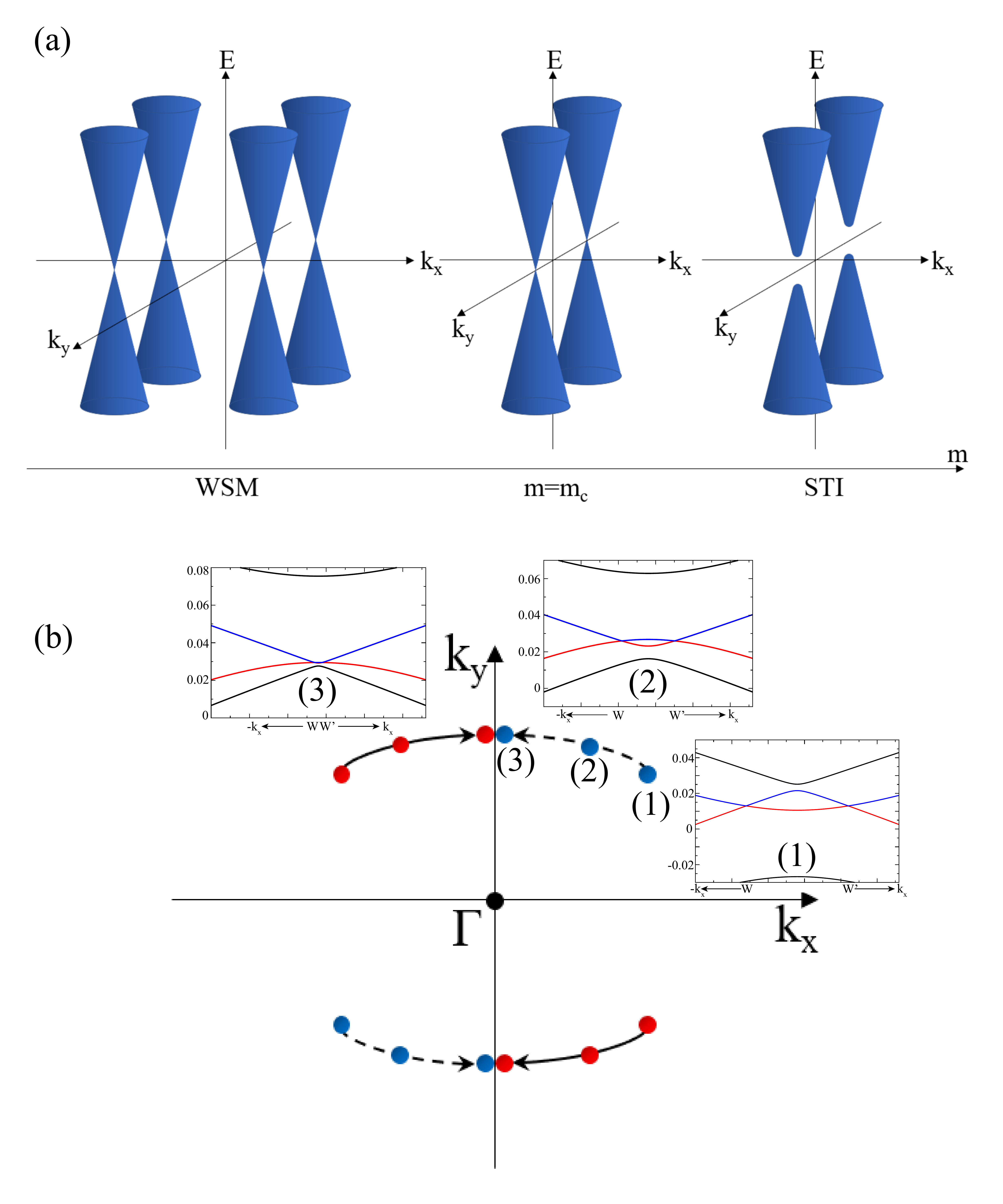}
\caption{(a) Schematic evolution of the topological phase transition from Cu$_{2}$SnS$_{3}$ (WSM) to Cu$_{2}$SnSe$_{3}$ (STI). (b) Band structures of Cu$_{2}$SnS$_{3-x}$Se$_{x}$ calculated by linear mixing of Wannier Hamiltonians with Se doping of 0.0 (1), 0.3 (2), and 0.365 (3). The $k$-path passes through the two Weyl points with opposite chirality along the $k_x$-axis. \label{fig6}}
\end{figure}

This work presents a simple and ideal model material system for realizing the TPT proposed by S. Murakami~\cite{murakami2011gap}. The process of Weyl nodes annihilation in pair and the TPT from a WSM to a TI are shown in Fig. \ref{fig6}(a). This is further simulated by using the linear mixing of the tight-binding Hamiltonians of Cu$_{2}$SnS$_{3}$ and Cu$_{2}$SnSe$_{3}$ constructed from the generated Wannier functions. We find that the Weyl points gradually approach the $k_x$ = 0 plane along the trajectory when the Se doping ratio increases as shown in Fig. \ref{fig6}(b). According to the calculation results, we find that at about S:Se = 0.63:0.37 the Weyl points finally annihilate on the $k_x$ = 0 plane and TPT from a WSM to a TI is realized with the MCN = 0 for $M_x$. The critical value of S:Se determined from linear mixing of tight-binding Hamiltonian is nearly the same as the first-principles calculation from VCA.

In order to further understand the TPT, we construct a two-band $k \cdot p$ model, which describes the bands around the crossing point between nodal-ring and $k_z$ axis in case without SOC. According to the band representations at this crossing point, the Hamiltonian expanded around it with momentum $q (q_x, q_y, q_z)$ can be simply written as:
\[H_0(q) = q_x \tau_x + q^2_y \tau_z + q_z \tau_z\],
where $q_x$, $q_y$ and $q_z$ are along $k_x$, $k_y$ and $k_z$ axis, respectively. In the absence of SOC, the Hamiltonian has linear dispersion along the $q_x$ and $q_z$-axis, but quadratic one along the $q_y$-axis. 

When SOC is further included, the two-band $k$ $\cdot$ $p$ model should become a four-band model because of the spin degree of freedom. At the zero point of $q$, the generators of the little group have $C_{2z}$, $M_x$ and $M_y$ symmetries. Their matrix representations can be obtained from the results of first-principles calculations:
\[M_x = i\tau_z \otimes s_x,\]
\[M_y = i\tau_0 \otimes s_y,\]
\[C_{2z} = i\tau_z \otimes s_z.\]

When SOC is considered, the spin components should be added to the previous model, which is equivalent to adding new mass terms to the original Hamiltonian. The mass term must commute with $C_{2z}$, $M_x$ and $M_y$. So the new Hamiltonian is given by:
\[H(q) = q_x \tau_x + q^2_y \tau_z + q_z \tau_z + m\tau_y s_y\]

For states in the $M_x$ plane, applying operation $M_x$ to $H(q_x = 0)$, we can get the block diagonal matrix of $H(q_y, q_z)$ in the $i$ or -$i$ eigenvalue subspaces of $M_x$~\cite{hsieh2012topological,PhysRevB.90.081112}:
\begin{equation}
UM_xU^{-1} =  \left(\begin{array}{cccc}
i& 0 & 0 & 0\\
0 & i& 0 & 0\\
0 & 0 & -i & 0\\
0 & 0 & 0 & -i\\
\end{array}\right)
\end{equation}

\begin{equation}
UH(q_y, q_z)U^{-1}= \left(\begin{array}{cccc}
-q^2_y-q_z & -m & 0 & 0\\
-m & q^2_y + q_z & 0 & 0\\
0 & 0 & -q^2_y - q_z & m\\
0 & 0 & m & q^2_y + q_z\\
\end{array}\right)
\end{equation}
The  subspace Hamiltonian of  $\pm i$ eigenvalue is
\[H^{\pm i}_{yz}(q) = \textbf{d} \cdot \textbf{$\sigma$} = \mp m\sigma_x - (q^2_y+ q_z)\sigma_z,\]
and the MCN for $M_x$ plane is
\[C^{\pm i}_{M_x}=\frac{1}{4\pi} \int dq_y dq_z \widehat{\textbf{d}} \cdot (\partial_{q_y} \widehat{\textbf{d}} \times \partial_{q_z} \widehat{\textbf{d}} ) = 0.\] 
Obviously, in $i$ or -$i$ subspace,  MCN = 0 on $M_x$ plane, which means that the $H_{yz}(q)$ is trivial.

However, for $M_y$ plane, we can obtain the nontrivial matrix of $H(q_x, q_z)$ by applying $M_y$ to $H(q_y = 0)$. 
\begin{small}
\begin{equation}
U'H(q_x, q_z)U'^{-1}= \left(\begin{array}{cccc}
-q_z & q_x+im & 0 & 0\\
q_x-im & q_z & 0 & 0\\
0 & 0 & -q_z & q_x-im\\
0 & 0 & q_x+im &q_z\\
\end{array}\right)
\end{equation}
\end{small}

The  subspace Hamiltonian of  $\pm i$ eigenvalue is
\[H^{\pm i}_{xz}(q) = \textbf{d} \cdot \textbf{$\sigma$} = q_x\sigma_x \mp m\sigma_y - q_z\sigma_z,\]
and the MCN for $M_y$ is
\[C^{\pm i}_{M_y}=\frac{1}{4\pi} \int dq_x dq_z \widehat{\textbf{d}} \cdot (\partial_{q_x} \widehat{\textbf{d}} \times \partial_{q_z} \widehat{\textbf{d}} ) = \mp \frac{1}{2}sgn(m).\] 
Since there are two crossing points on the $M_y$ plane because of the TRS without SOC, the mirror Chern number is
\[C_{M_y} = \left| C^{\pm i}_{M_y} \right| \times 2 = 1.\]
These results are consistent with those of Cu$_{2}$SnS$_{3}$ from first-principles calculations.

\section{Summary}  
Through first-principles calculations, we have proposed that Cu$_{2}$SnS$_{3}$ and Cu$_{2}$SnSe$_{3}$ can be used to model the topological phase transition from a WSM to a TI. In the absence of SOC, both of them are the simplest nodal line semimetal with only a single nodal ring centering at $\Gamma$, which is protected by $M_x$ symmetry and lies in the mirror plane. When SOC is taken into account, they are quite different. For Cu$_{2}$SnS$_{3}$, the nodal-ring evolves into two pairs of Weyl points in the $k_z$ = 0 plane, as indicated by the different MCN for $M_x$ and $M_y$ planes, namely MCN = 0 for $M_x$ and MCN = 1 for $M_y$ plane. For Cu$_{2}$SnSe$_{3}$, the nodal-ring is fully gapped and the system becomes a strong TI, as indicated by the same MCN = 1 for both $M_x$ and $M_y$ planes. The difference in them comes from the different strength of effective SOC which can be systematically tuned by doping Se into Cu$_{2}$SnS$_{3}$. Employing VCA, we have simulated the doping concentration continuously to show the movement of Weyl points and their annihilation in the $M_x$ plane during the TPT. The critical doping level is S:Se = 0.63:0.37. We have also constructed a $k \cdot p$ model to explain these results. Here, it must be noted that all the above results on specific materials are based on the GGA calculations, which usually overestimates the band inversion. The previous work~\cite{choi2015optical,mesbahi2017dft,ZHOU201877} mentioned that Cu$_{2}$SnS$_{3}$ and Cu$_{2}$SnSe$_{3}$ are gapped insulators in reality and our improved hybrid functional (HSE06) calculations shown in Appendix are consistent with them. There are still some family compounds, like Cu$_{2}$SiTe$_{3}$, Cu$_{2}$GeSe$_{3}$, Cu$_{2}$GeTe$_{3}$, and Cu$_{2}$SnTe$_{3}$, keeping the band inversion and their band topology can be analyzed similarly. Nevertheless, our work is of importance and usefulness for theoretically studying the topological states and phase transitions among them.

\section{Acknowledgement} 
We acknowledge the supports from the National Natural Science Foundation (Grant No. 11925408, 11974076, 11921004 and 12188101), the Ministry of Science and Technology of China (Grant No. 2018YFA0305700), the Chinese Academy of Sciences (Grant No. XDB33000000), the K. C. Wong Education Foundation (GJTD-2018-01), the Beijing Natural Science Foundation (Z180008), and the Beijing Municipal Science and Technology Commission (Z191100007219013). CY is supported by the Swiss National Science Foundation (200021-196966). WZ is supported by the Key Project of Natural Science Foundation of Fujian Province (2021J02012).

\newpage{}
\begin{widetext}
\appendix

\maketitle
\section{Band structures of the other members in Cu$_{2}$SnS$_{3}$ family}
In this section, we present the bulk band structures of the other members in Cu$_{2}$SnS$_{3}$ family without and with spin-orbit coupling (SOC), which are all calculated by GGA.
\begin{figure}[htbp!]
\centering
\includegraphics[width=1.0\textwidth]{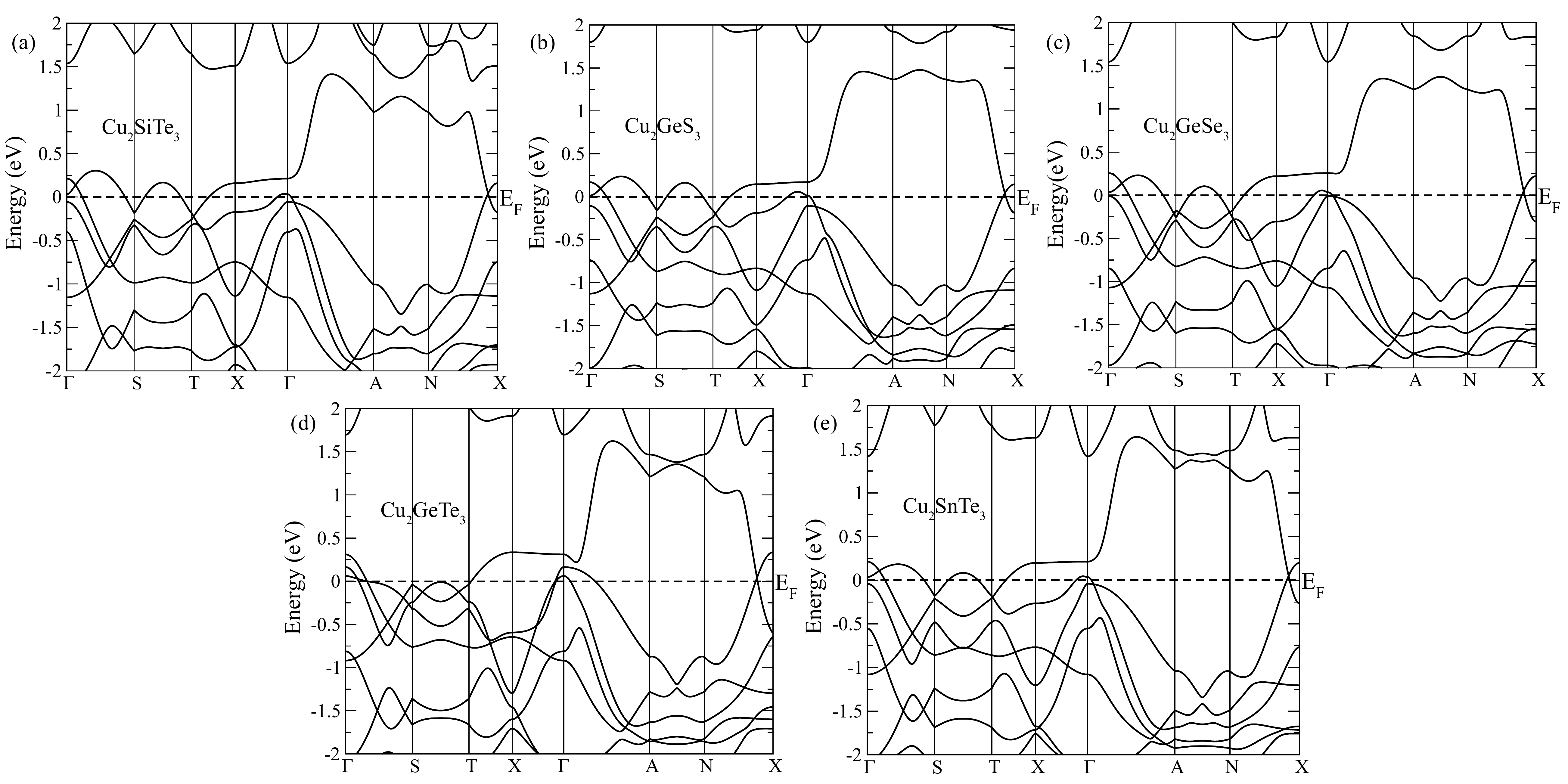}
\caption{Band structures of Cu$_{2}$SiTe$_{3}$, Cu$_{2}$GeS$_{3}$, Cu$_{2}$GeSe$_{3}$, Cu$_{2}$GeTe$_{3}$ and Cu$_{2}$SnTe$_{3}$ along high symmetry points without the spin-orbit coupling. \label{figs1}}
\end{figure}

\begin{figure}[htbp!]
\centering
\includegraphics[width=1.0\textwidth]{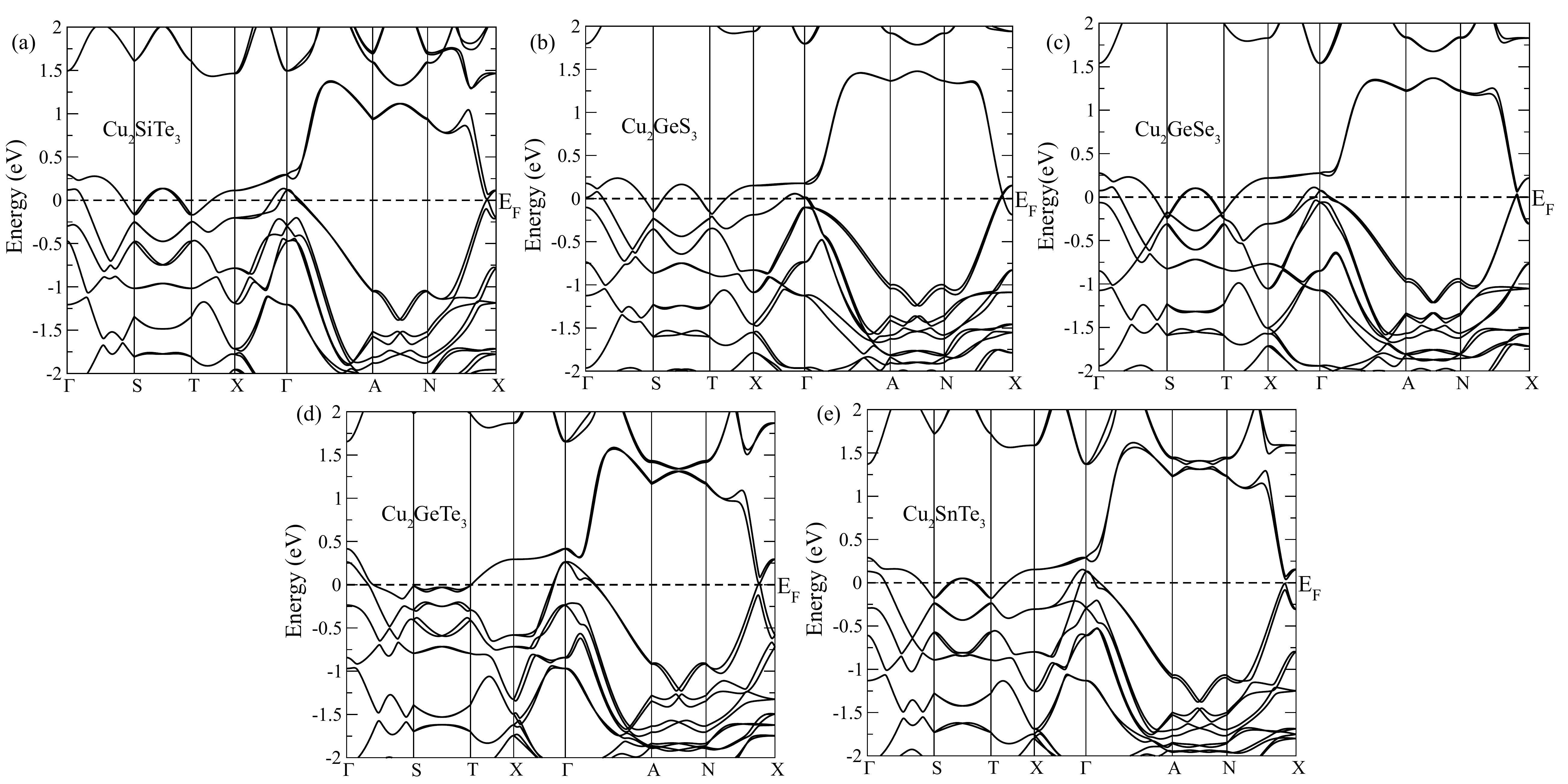}
\caption{Band structures of Cu$_{2}$SiTe$_{3}$, Cu$_{2}$GeS$_{3}$, Cu$_{2}$GeSe$_{3}$, Cu$_{2}$GeTe$_{3}$ and Cu$_{2}$SnTe$_{3}$ along high symmetry points with the spin-orbit coupling. \label{figs2}}
\end{figure}

\section{MCNs of $M_x$ and $M_y$ planes for other materials of Cu$_{2}$SnS$_{3}$ family}
In this section, we give the MCNs of $M_x$ and $M_y$ planes for other materials of Cu$_{2}$SnS$_{3}$ family. It's obvious to see that all the members have MCN = 1 for $M_y$ plane, but for $M_x$ plane, except for Cu$_{2}$SnS$_{3}$, all the other members have MCN = 1. This is a very novel phenomenon.

\begin{figure}[htbp!]
\centering
\includegraphics[width=1.0\textwidth]{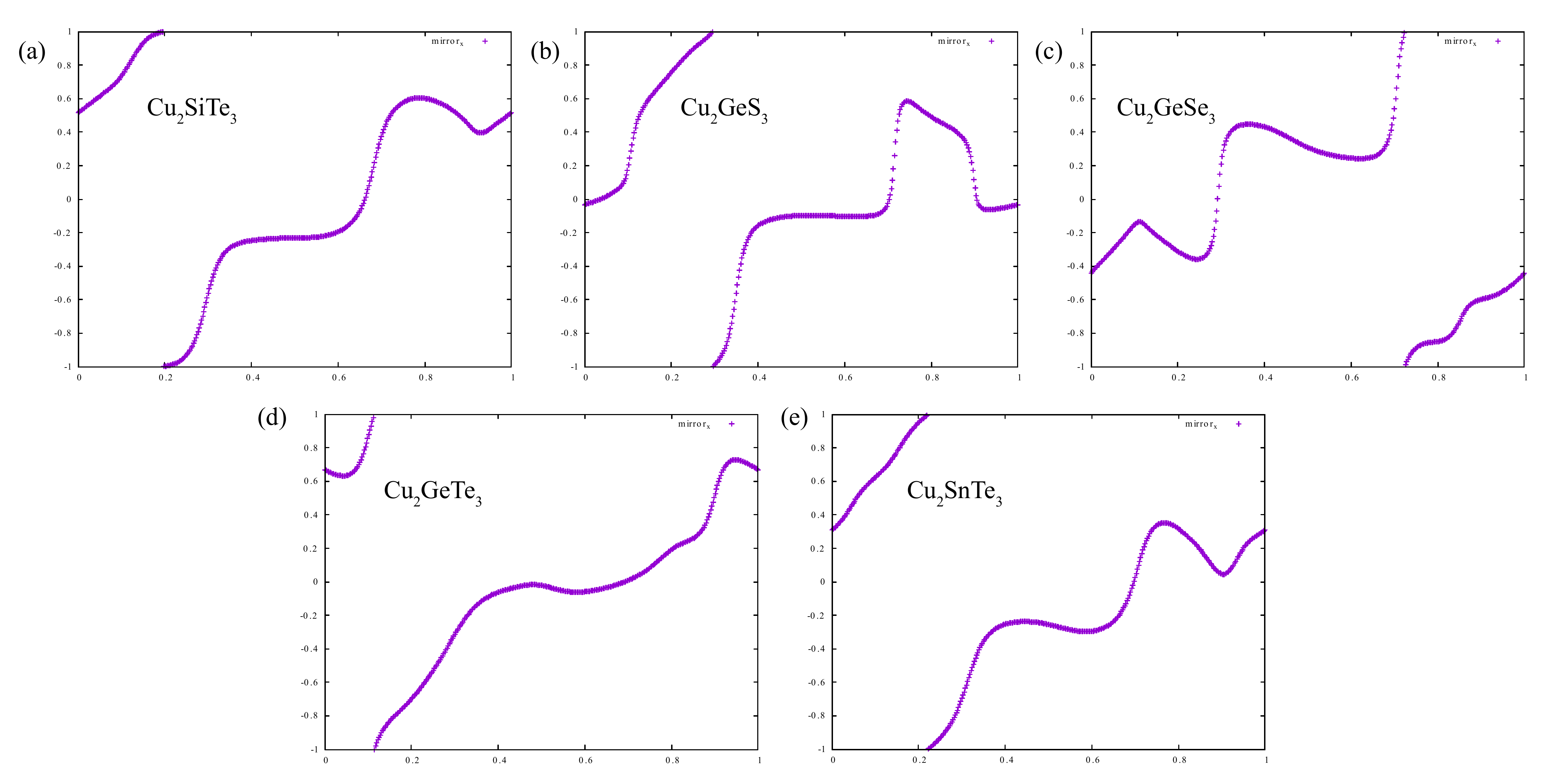}
\caption{Mirror Chern numbers (MCN) of Cu$_{2}$SiTe$_{3}$, Cu$_{2}$GeS$_{3}$, Cu$_{2}$GeSe$_{3}$, Cu$_{2}$GeTe$_{3}$ and Cu$_{2}$SnTe$_{3}$ for $M_x$ plane. \label{figs3}}
\end{figure}

\begin{figure}[htbp!]
\centering
\includegraphics[width=1.0\textwidth]{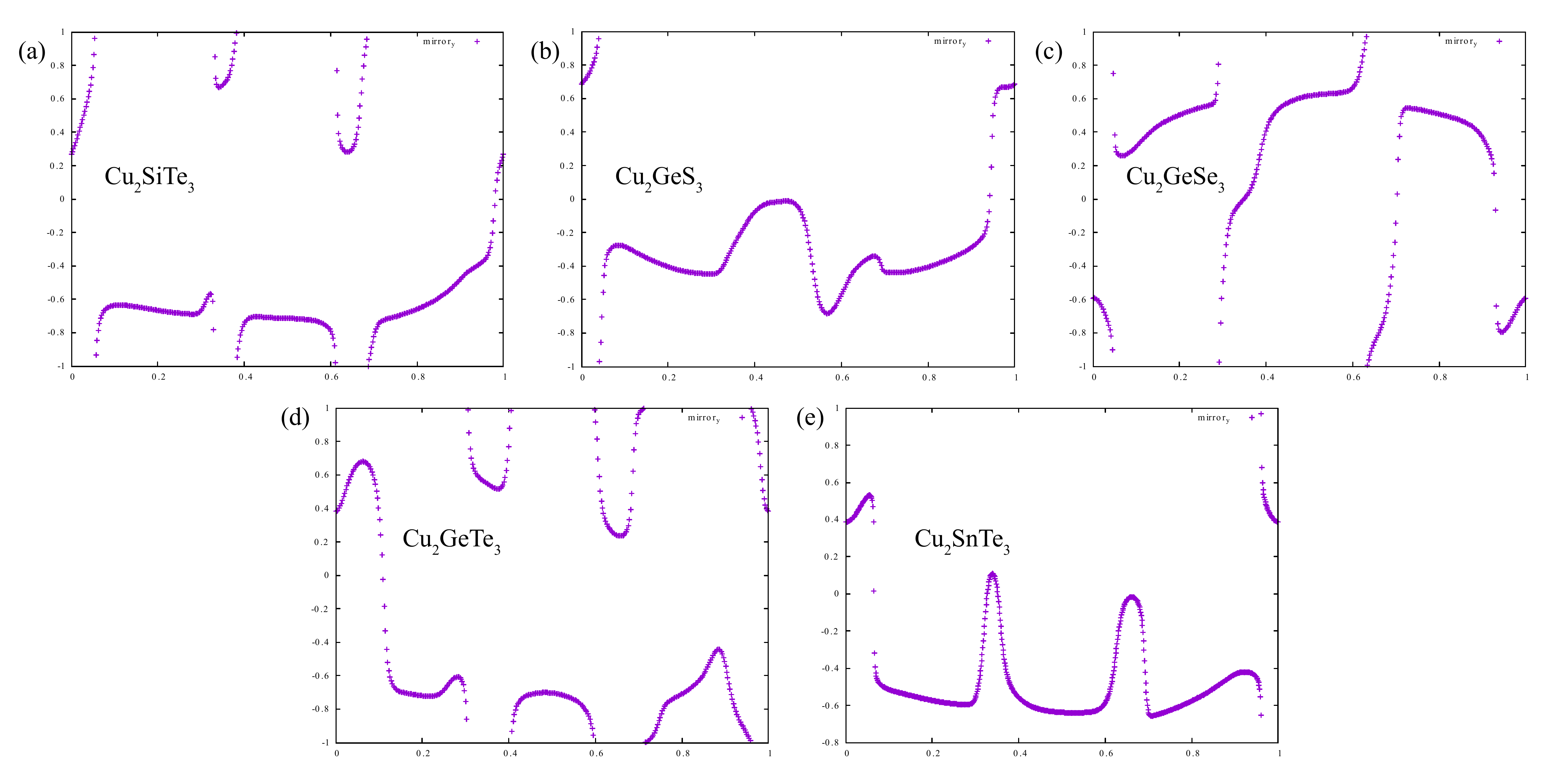}
\caption{Mirror Chern numbers (MCN) of Cu$_{2}$SiTe$_{3}$, Cu$_{2}$GeS$_{3}$, Cu$_{2}$GeSe$_{3}$, Cu$_{2}$GeTe$_{3}$ and Cu$_{2}$SnTe$_{3}$ for $M_y$ plane.\label{figs4}}
\end{figure}

\section{The distribution of Weyl points in Cu$_{2}$SnS$_{3}$ family}
In this section, we give the position of Weyl points for these materials of Cu$_{2}$SnS$_{3}$ family.

\begin{table}[H]
\centering
\caption{The distribution of Weyl points in Cu$_{2}$SnS$_{3}$ family.}
\setlength{\tabcolsep}{4mm}{
\begin{tabular}{ccccccc}
material&a ({\AA})&b ({\AA})&c ({\AA})&Weyl points ($k_x$, $k_y$, $k_z$) (2$\pi$/a)&energy (eV)&number\\
\hline
Cu$_{2}$SiTe$_{3}$&4.2527&12.5882&5.9446&upper and lower surfaces of BZ& &8 small nodal-rings\\
Cu$_{2}$GeS$_{3}$&3.7660&11.3210&5.2100&(0.0510, 0.4044, 0.3047)&-0.0749&8\\
 & & & &(0.0287, 0.2935, 0.2154)&0.0062&8\\
Cu$_{2}$GeSe$_{3}$&3.9600&11.8600&5.4850&(0.0617, 0.3173, 0.2536)&-0.0111&8\\
Cu$_{2}$GeTe$_{3}$&4.2115&12.6410&5.9261&(0.1313, 0.0775, 0.1327)&0.1290&8\\
 & & & &upper and lower surfaces of BZ& &4 large nodal-rings\\
Cu$_{2}$SnS$_{3}$&3.8937&11.5720&5.4436&(0.0061, 0.1120, 0.0000)&0.0014&4\\
Cu$_{2}$SnSe$_{3}$&4.1158&12.2715&5.7528&Topological insulator (TI)& &0\\
Cu$_{2}$SnTe$_{3}$&4.2740&12.8330&6.0430&(0.0576, 0.1502, 0.1226)&0.0748&8\\
 & & & &(0.0782, 0.4763, 0.3110)&-0.1824&8\\
\hline
\end{tabular}}
\label{weyl}
\end{table}

\begin{figure}[htbp!]
\centering
\includegraphics[width=1.0\textwidth]{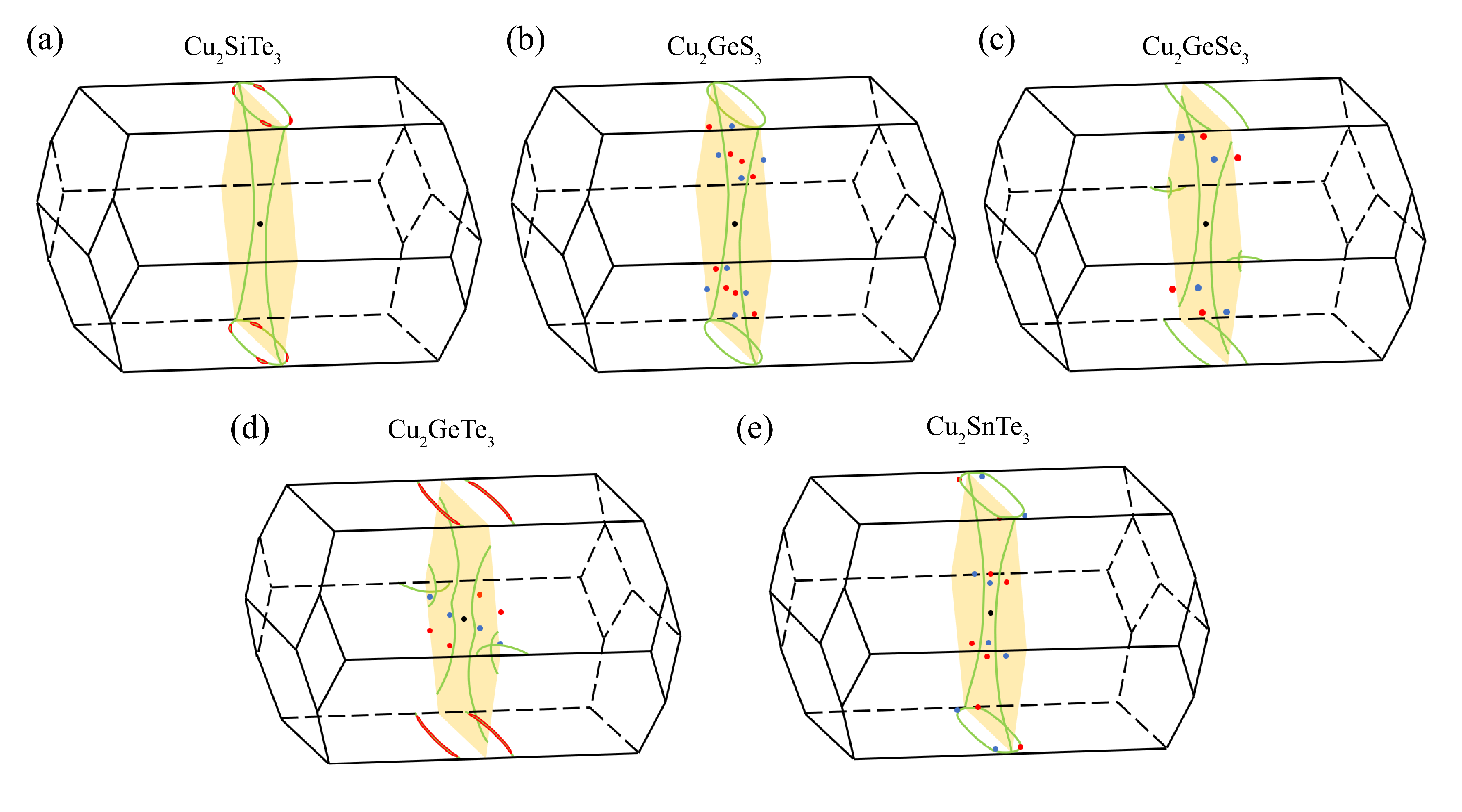}
\caption{The schematic diagrams of Weyl points distribution of Cu$_{2}$SiTe$_{3}$, Cu$_{2}$GeS$_{3}$, Cu$_{2}$GeSe$_{3}$, Cu$_{2}$GeTe$_{3}$, and Cu$_{2}$SnTe$_{3}$ (The green line represents the nodal-chain and nodal-ring. The blue dots denote the Weyl points with negative chirality and the red dots denote the Weyl points with positive chirality). \label{figs6}}
\end{figure}

\section{Band structures of Cu$_{2}$SnS$_{3}$ family with HSE06}
In this section, we show the bands calculation results with HSE06 for all materials without SOC and partial materials with SOC.

\begin{figure}[htbp!]
\centering
\includegraphics[width=1.0\textwidth]{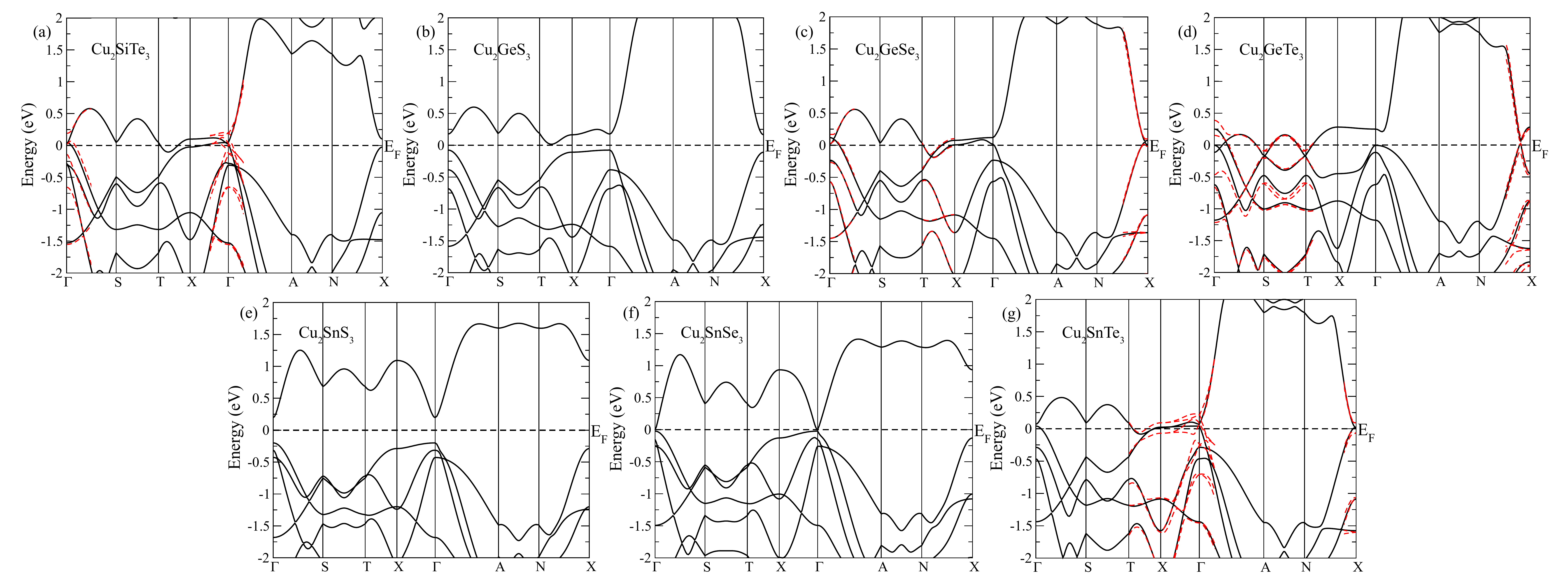}
\caption{The band structures of Cu$_{2}$SiTe$_{3}$, Cu$_{2}$GeS$_{3}$, Cu$_{2}$GeSe$_{3}$, Cu$_{2}$GeTe$_{3}$, Cu$_{2}$SnS$_{3}$, Cu$_{2}$SnSe$_{3}$ and Cu$_{2}$SnTe$_{3}$ with HSE06 along high symmetry points without the spin-orbit coupling (The red dotted lines in part of the figures are the bands with SOC for comparison). \label{figs7}}
\end{figure}
\end{widetext}

\newpage{}
\bibliography{ref}

\end{document}